\documentclass[a4paper, 10pt]{article}
\usepackage{amsmath,bbm,mathrsfs}
\usepackage{graphicx}
\usepackage{amsfonts}
\usepackage{amssymb}
\usepackage{ae}
\usepackage{dsfont}

\usepackage[T1]{fontenc}
\usepackage{amssymb, amsthm, amscd}
\usepackage{amsmath}
\usepackage{graphicx}

\usepackage{multirow}
\usepackage{ctable}
\usepackage{bigstrut}

\usepackage{fullpage}

\def\textbf#1{{\bf #1}}
\def\be{\begin{equation}}
\def\ee{\end{equation}}
\def\ben{\begin{eqnarray}}
\def\een{\end{eqnarray}}
\def\eea{\end{array}}
\def\bea{\begin{array}}

\newcommand{\tr}[1]{\mathrm{tr}#1}
\newcommand{\bi}{\begin{itemize}}
\newcommand{\ei}{\end{itemize}}
\newcommand{\ket}[1]{|#1\rangle}
\newcommand{\bra}[1]{\langle#1|}

\newcommand{\proj}[1]{\ket{#1}\!\bra{#1}}

\newcommand{\Ke}[1]{\big|#1\big\rangle}
\newcommand{\Br}[1]{\big< #1\big|}

\newcommand{\ke}[1]{|#1\rangle}

\DeclareMathOperator{\identity}{\mathbb{I}}

\newtheorem{theorem}{Theorem}

\begin{document}

\newcommand{\eg}{{\it{e.g.~}}}
\newcommand{\ie}{{\it{i.e.~}}}
\newcommand{\etal}{{\it{et al.}}}

\title{Can bipartite classical information resources be activated?}
\author{Giuseppe Prettico\footnote{giuseppe.prettico@icfo.es}\\ \normalsize{\it ICFO-Institut de Ciencies Fotoniques, 08860 Castelldefels, Barcelona, Spain}\\
\and Antonio Ac\'in \\\normalsize{\it ICFO-Institut de Ciencies
Fotoniques, 08860 Castelldefels, Barcelona, Spain}
\\\normalsize{\it ICREA-Instituci\'o Catalana de Recerca i Estudis Avan\c{c}ats,
08010 Barcelona, Spain}} \maketitle

\section*{\normalsize{ Abstract}}
Non-additivity is one of the distinctive traits of Quantum
Information Theory: the combined use of quantum objects may be
more advantageous than the sum of their individual uses.
Non-additivity effects have been proven, for example, for quantum
channel capacities, entanglement distillation or state estimation.
In this work, we consider whether non-additivity effects can be
found in Classical Information Theory. We work in the secret-key
agreement scenario in which two honest parties, having access to
correlated classical data that are also correlated to an
eavesdropper, aim at distilling a secret key. Exploiting the
analogies between the entanglement and the secret-key agreement
scenario, we provide some evidence that the secret-key rate may be
a non-additive quantity. In particular, we show that correlations
with conjectured bound information become secret-key distillable
when combined. Our results constitute a new instance of the subtle
relation between the entanglement and secret-key agreement
scenario.

\section*{\normalsize{ Introduction}}
Classical communication systems are governed by classical
information theory, a vast discipline whose birth coincides with a
seminal paper of Claude Shannon~\cite{Shannon}. Among his
contributions, Shannon introduced the concept of channel capacity,
which quantifies the maximum communication rate that can be
achieved over a classical channel. One key feature of the channel
capacity is its additivity: the total capacity of several channels
used in parallel is simply given by the sum of their individual
capacities. This fact implies thus that the channel capacity
completely specifies channel's ability to convey classical
information.

Moving to the quantum domain, the quantum channel capacity
captures the ability of a quantum channel to transmit quantum
information. Smith and Yard~\cite{SmYa} proved recently that the
quantum capacity is not additive. In particular, they provide
examples of two channels with zero quantum capacity that define a
channel with strictly positive quantum capacity when combined.
This intriguing quantum effect is known as activation and can
generally be understood as follows: the combined use of quantum
objects can be more advantageous than the sum of their individual
uses. In the last years, an intense effort has been devoted to the
study of non-additivity effects in Quantum Information Theory.
Classical and private communication capacity of quantum channels
were later shown not to be additive in
Refs~\cite{hastings,private}. Nowadays, non-additivity is
considered to be one of the distinctive traits of Quantum
Information Theory.

Before the results by Smith and Yard, however, non-additivity
effects had also been observed in Entanglement Theory in the
context of entanglement distillation. There, one is interested in
the problem of whether pure-state entanglement --pure entanglement
in what follows-- can be extracted from a given state shared by
several observers using local operations and classical communication
(LOCC). In Ref.~\cite{SuperActiv}, the authors provide examples of
multipartite states that (i) are non-distillable (bound) when
considered separately but (ii) define a distillable state when
taken together. Moving to the case of two parties,
and leaving aside activation-like results as those of~\cite{H3},
it remains unproven whether entangled states can be activated.
There is however some evidence of the existence of pairs of bound
(non-distillable) entangled states that give a distillable state
when combined~\cite{SST,Activ}.

In this work we are interested in the question of whether
non-additivity effects can be observed in Classical Information
Theory. As mentioned, classical channel capacities are known to be
additive. Therefore, we move our considerations to distillation
scenarios. In particular, we focus on the classical secret-key
agreement scenario in which two honest parties, having access to
correlated random variables, also correlated with an
adversary, aim at establishing a secret key by local operations
and public communication (LOPC). While the activation of classical
resources has been shown in a multipartite key-agreement scenario
in~\cite{ref:multibouninf,ref:superactivboundinf}, here we
consider the more natural case of two honest parties. In our
study, we exploit the analogies between the secret-key agreement
and entanglement scenario noted in~\cite{ref:linking}. Based on
the results of~\cite{Activ}, we provide evidence that activation
effects may be possible in the completely classical bipartite
key-agrement scenario. Our findings, therefore, suggest that the
classical secret-key rate is non-additive.

This article is structured as follows: Section~\ref{concepts}
contains a brief introduction to the entanglement and the
secret-key agreement scenario. After pointing out the analogies
between the two scenarios, our main results are derived in
section~\ref{Results}. Section~\ref{Conclusions} concludes with a
discussion of how our findings are related to other results and
conjectures in the field.

\section{Entanglement vs secret correlations}
\label{concepts}

The aim of this section is to introduce the entanglement and
secret-key agreement scenario. As first noted in~\cite{Linking},
there are several analogies between these two scenarios despite
the fact that they involve objects of different nature, namely
entangled quantum states vs classical joint probability
distributions. These analogies play a key role in the derivation
of our results in the next section.

\subsection{Entanglement scenario}\label{QuanScen} A
maximally entangled state of two qubits represents the most
representative example of a bipartite entangled state and is an
essential ingredient in many applications of quantum information
theory~\cite{Ben}. It is defined as:
\begin{equation}\label{ebit}
\ke\phi=\frac{1}{\sqrt{2}}\left(\Ke {00}+\Ke {11}\right)_{AB}
\end{equation}
The relevance of this state for communication purposes is due
essentially to two main facts: first, for each projective
measurement by one of the observers, there exists another
measurement by the other observer giving perfectly correlated
results. Second, being a pure state, no third party can be
correlated with it. State~\eqref{ebit} represents the basic unit
of entanglement and is also known as \emph{ebit}, for entangled
bit. This is because an asymptotically large number of copies of
an arbitrary pure entangled state can be converted into another
asymptotically large number of ebits in a reversible
way~\cite{entrev}.

In any realistic situation, quantum states are affected by noise.
In the case of a composite system shared by two observers, Alice
and Bob (also $A$ and $B$), the ideal pure entangled states is
mapped into a mixed state $\rho_{AB}$. Any noise can be modeled as
interaction with an environment, $E$, as any state can be seen as
the trace of a pure state on a sufficiently large environment. In
the bipartite case considered here one has
$\rho_{AB}=\tr_E\proj{\psi_{ABE}}$. Of course, the interaction
with the environment deteriorates the entanglement present in the
state.
Thus, given a generic quantum state $\rho_{AB}$, or equivalently
the whole tripartite state $\ket{\psi_{ABE}}$, quantifying the
entanglement between $A$ and $B$ is a fundamental question. Two
quantifiers play a crucial role because of their operational
meaning: the entanglement cost and the entanglement of
distillation. Both quantities are defined in the asymptotic
scenario consisting of an asymptotically large number of identical
copies of the state. The entanglement cost~\cite{ref:entcost},
denoted by $E_{c}$, quantifies the number of ebits per copy needed
for the formation of the given quantum state by LOCC. The
entanglement of distillation~\cite{ref:distillent}, denoted by
$E_{D}$, indicates the amount of ebits per copy that can be
obtained from it by LOCC. For a state $\rho_{AB}$,
$E_{c}(\rho_{AB}) > 0$ implies that the state is entangled, while
$E_{D}(\rho_{AB}) > 0$ indicates that some pure entanglement can
be extracted from it. Clearly, it holds that $E_{c} \geq E_{D}$,
as one cannot extract from a state more entanglement than needed
for its preparation. Interestingly, there are states that display
an intriguing form of irreversibility: despite having a positive
entanglement cost ($E_{c}>0$), they are non-distillable ($E_{D}
=0$). These states are called \emph{bound
entangled}~\cite{ref:boundent}. Consequently, the whole set of
entangled states is composed of distillable, or free entangled
states, and bound entangled states.

Detecting whether a given state is non-distillable is in principle
a very hard question, as one has to prove that no LOCC protocol
acting on an arbitrary number of copies of the state is able to
extract any pure entanglement. However, a very useful result
derived in~\cite{ref:boundent} shows that a quantum state that
remains Positive under Partial Transposition \cite{Peres} (PPT) is
non-distillable. Whether Non-Positivity of the Partial
Transposition, or Negative Partial Transposition (NPT), is
sufficient for entanglement distillability is probably the main
open question at the moment in Entanglement Theory.
Evidence~\cite{DCLB,NPPTConj} has been given for the existence of
NPT states that are bound entangled (see however~\cite{watrous}).
Note that the existence of these states would imply that the set
of non-distillable states is not convex and that entanglement of
distillation is non-additive~\cite{SST}. A necessary and
sufficient condition for the distillability of a quantum state is
provided by the following
\begin{theorem}
\label{Projectors} A state $\rho$ acting on ${\cal H}={\cal H}_A
\otimes {\cal H}_B$ is distillable if and only if there exist a
finite integer number $n\geq 1$ and two dimensional projectors
$P:{\cal H}^{\otimes n}_A \rightarrow \mathbb C^2 $ and $Q:{\cal
H}^{\otimes n}_B \rightarrow \mathbb C^2 $ such that the state
\begin{align}
\label{pqpq}
\rho'=(P\otimes Q)\rho^{\otimes n} (P\otimes Q)^{\dagger}
\end{align}
is entangled.
\end{theorem}
\noindent Actually, since the resulting state acts on $ \mathbb
C^2\otimes \mathbb C^2$, this is equivalent to demand that $\rho'$
is NPT, as this condition is necessary and sufficient for
entanglement in the two-qubit case~\cite{horopla}. Furthermore, it
is worth mentioning here that, if such a projector exists for some
number $k$ of copies, the state is said to be $k-distillable$.

\subsection{Secret-key agreement
scenario}\label{ClassScen}

The main scope of this section is to introduce the secret-key
agreement scenario. This scenario consists of two honest parties,
again Alice and Bob, who have access to correlated information,
described by two random variables $X$ and $Y$. These variables are
also correlated to a third random variable $Z$ that belongs to an
adversarial party, the eavesdropper Eve, denoted by $E$. All the
correlations among the three parties are described by the
probability distribution $P(XYZ)$. The honest parties aim at
mapping the initial correlations into a secret key by LOPC, which
is the natural set of operations between the honest parties.

Similar questions as above can be addressed in this completely
classical scenario. The classical equivalent of a maximally
entangled state is a secret bit. $A$ and $B$ share perfect secret
bits whenever $P(XYZ)$ is such that the eavesdropper is factored
out, $P(XY)\times P(Z)$, and their variables can take two possible
values, $X,Y=0,1$, that are perfectly correlated and random,
$P(X=Y=0)=P(X=Y=1)=1/2$. Similarly as above, given some initial
correlations, the goal is to quantify its secrecy content. The
classical analog of $E_{c}$ is the information of formation,
denoted by $I_{f}$~\cite{ref:boundinfo}. It is said that the
probability distribution $P(XYZ)$ contains secret correlations (or
secret bits) whenever $I_{f}(P(XYZ))>0$. For distillation, the
natural classical analog is the secret-key rate
\cite{ref:keyrate}, denoted by $S(X:Y \| Z)$, which quantifies the
number of secret bits that can be distilled from given
correlations by LOPC. Due to the difficulty of computing the
previous quantities, it is useful to establish bounds on them. The
intrinsic information~\cite{ref:keyrate}, $I(X;Y \downarrow Z)$,
provides a lower bound to the information of
formation~\cite{ref:boundinfo} and an upper bound to the
secret-key rate \cite{ref:keyrate}:
\begin{equation}\label{eq:BoundIntinsic}
S(X;Y\|Z) \leq I(X;Y \downarrow Z) \leq {I_{f}(X;Y|Z)}
\end{equation}
It is defined as the minimal mutual information between $A$ and
$B$ conditioned on $E$ over all possible maps $Z \rightarrow
\bar{Z}$ the eavesdropper can perform, that is,
\begin{equation}\label{eq:Intinsic}
I(X;Y \downarrow Z):=\min_{P_{\overline{Z}|Z}}\left [
{I(X;Y|\overline{Z}): P_{XY\overline{Z}}=\sum_z P_{XYZ}\cdot
P_{\overline{Z}|Z}} \right ]
\end{equation}
In Ref~\cite{ChReWo} it was shown that it is sufficient to consider the output alphabet
$\bar{Z}$ of the same size as the input alphabet $Z$.

A main open question in this scenario is whether there
exist non-distillable secret correlations with strictly positive
information of formation. These correlations are named \emph{bound
information}, as they would constitute a classical cryptographic
analog of bound entanglement~\cite{Linking}. Compared to the
entanglement scenario, identifying a single example of
non-distillable correlations is much harder, due to the lack of a
simple mathematical criterion, as Partial Transposition, to detect
it. In a multipartite scenario, say of three honest parties plus
an eavesdropper, the possibility of splitting the honest parties
into different bipartitions hugely simplifies the problem and,
indeed, there are examples of correlations that require secret
bits for the preparation and from which no secret bits can be
extracted~\cite{ref:multibouninf}. The problem remains open for
two honest parties, although evidence has been provided for the
existence of bound information~\cite{Linking}.

When studying the distillation properties of some given
correlations, one usually employs \textit{Advantage Distillation
(AD)} protocols. These protocols were first introduced by
Maurer~\cite{Maurer} to show how two honest parties may be able to
extract a secret key even in cases in which Bob has less
information than Eve about Alice's symbols. Crucial to achieve
this task is feedback, that is, \textit{two way communication}
between the honest parties. The general structure of an AD
protocol is as follows \cite{AGS} (without loss of generality we
assume that Alice's and Bob's variables have the same size $d$):
Alice first generates randomly a value $\zeta$. She chooses a
vector of $N$ symbols from her string of data,
$\textbf{a}=(a_1,\ldots,a_N)$, and publicly announces their
positions to Bob. Later she sends him the $N$-dimensional vector
$\bar{\textbf{a}}$ whose components $\bar{a} _k$ are such that
$a_k \oplus \bar{a} _k = \zeta$ holds $\forall k$. Here, $\oplus$
is the sum modulo $d$. Bob sums $\bar{\textbf{a}}$ to his
corresponding symbols. If he obtains always the same value $\chi$,
then he accepts (this means that with very high probability
$\chi=\zeta$) otherwise both discard the $N$ symbols. Although its
yield is very low with increasing $N$, AD protocols allow the
honest parties to distill a key even in a priori disadvantageous
situations in which Eve has more information than Bob on Alice's
symbols. Such protocols are used in what follows to estimate the
distillability properties of correlations. Obviously, the fact
that we are unable to map some correlations into a secret key by
AD protocols does not mean that these correlations are
non-distillable. At best, it can be interpreted as some evidence
of bound information.

Finally, another concept used in the sequel is that of
\textit{binaryzation}, which can be understood as the classical
analog of the quantum projection onto 2-qubit subspaces used in
Theorem \ref{Projectors}. As in the quantum case, Alice and Bob
agree on two possible values, not necessarily the same, and
discard all instances in which their random variables take
different values. Then, they project their initial distribution
onto a smaller (and usually simpler) two-bit distribution.

\subsection{From Quantum States to Classsical
Probabilities}\label{QuanClass}

It is clear from the previous discussion that the entanglement and
secret-key agreement scenarios have a similar formulation. One can
go further and establish connections between the entanglement of
bipartite quantum states and the tripartite probability
distributions that can be derived from them~\cite{Linking}. Not
surprisingly, the transition from quantum states to classical
probabilities is through measurements (on the quantum states).
Note also that, while in the quantum case the state between Alice
and Bob also specifies the correlations with the environment,
possibly under control of the eavesdropper, in the classical
cryptographic scenario it is essential to define the correlations
with the eavesdropper for the problem to be meaningful.

As mentioned, if Alice and Bob share a state $\rho_{AB}$, the
natural way of including Eve is to assume that she owns a
purification of it. In this way the global state of the three
parties is a pure tripartite $\Ke{\psi_{ABE}}$ such that
$\rho_{AB}=\tr_E \left( \Ke {\psi_{ABE}} \Br {\psi_{ABE}} \right
)$. After this purification, measurements by the three parties,
$M_X$, $M_Y$ and $M_Z$, respectively, map the state into a
tripartite probability distribution:
\begin{equation}
\label{pxyz} P(XYZ)=\tr\left( M_X\otimes M_Y\otimes M_Z \ \Ke
{\psi_{ABE}}\Br {\psi_{ABE}} \right )
\end{equation}
It has been shown that (i) if the initial quantum state is
separable, there exists a measurement by the eavesdropper such
that the probability distribution~\eqref{pxyz} has zero intrinsic
information for all measurements by Alice and
Bob~\cite{Linking,curty} and also zero information of
formation~\cite{ref:SecCor} and (ii) if the initial state is
entangled, there exist measurements by Alice and Bob such that the
probability distributions~\eqref{pxyz} has strictly positive intrinsic
information for all measurements by Eve~\cite{ref:SecCor}.

Concerning the cryptographic classical analog of bound
entanglement, bound information was first conjectured in
Ref.~\cite{ref:linking}. There, local measurements were applied to
known examples of bound entangled states. It was then shown that
the resulting tripartite probability distributions have positive
intrinsic information but no known protocol allows the honest
parties to distill a secret key. Of course, this does not mean
that the distribution is non-distillable. Note however that the
existence, and activation, of bound information was proven in a
multipartite scenario consisting of three honest parties, plus the
eavesdropper, in Ref.~\cite{ref:multibouninf} (see
also~\cite{ref:superactivboundinf}). The examples of multipartite
bound information given in these works were derived from existing
multi-qubit bound entangled states.

\section{Is the secret-key rate a non-additive quantity?}
\label{Results}

This section presents our main results. Exploiting the analogies
between the entanglement and secret-key agreement scenarios, we
study whether it is possible to derive a cryptographic classical
analog of the activation of distillable entanglement between
bipartite quantum states given in Ref.~\cite{Activ}. This result
is reviewed in the following section. We then map the involved
quantum states onto probability distributions and study their
secrecy properties. After applying classical distillation
protocols, we show how the honest parties are able to distill a
secret key from each of the distributions for the same range of
parameters as in the quantum regime ($E_D > 0$). Finally, we
introduce a distillation protocol analogue to the one used for the
quantum activation. We prove that this protocol activates
probability distributions containing conjectured bound
information, although we cannot completely recover the quantum
region.
\subsection{Quantum Activation}
As mentioned, we start by presenting the example of activation of distillable
entanglement given in Ref.~\cite{Activ}. After
introducing the states involved in this example, we review their
distillability properties and the quantum protocol that attains
the activation.

\subsubsection{Quantum States}
States that are invariant under a group of symmetries play a
relevant role in the study of entanglement. The two classes of
symmetric states considered here are \textit{Werner}
states~\cite{Werner} and the \textit{symmetric} states of
Ref.~\cite{VollWer,Activ}, named in what follows symmetric states
for the sake of brevity.

\textsc{Werner States.}
Acting on an Hilbert space
$\mathcal{H}=\mathcal{H}_A\otimes \mathcal{H}_B$ with dimensions
$\textit{dim}(\mathcal{H}_A)=\textit{dim}(\mathcal{H}_B)=d$,
 and commuting with all unitaries $U\otimes U$, Werner states can be expressed as:

\begin{equation}\label{eq:wernerP}
\rho _W (p)=p \frac{\mathcal{A}_d }{tr(\mathcal{A}_d)}+(1-p) \frac{\mathcal{S}_d}{tr(\mathcal{S}_d)}
\end{equation}
where $\mathcal{A}_d=(\mathds{1}-\Pi
_d)/2,\quad\mathcal{S}_d=(\mathds{1}+\Pi _d)/2$ are the projector
operators onto the antisymmetric and symmetric subspaces,  $\Pi
_d$ is the flip operator and $tr(\mathcal{A}_d)=d(d-1)/2 $,
$tr(\mathcal{S}_d)=d(d+1)/2 $. It is known that states
(\ref{eq:wernerP}) are entangled and NPT iff $p>p_s=1/2$. Moreover
they are distillable, actually $1-distillable$, if
$p>p_{1d}=3\tau/(1+3\tau)$, where
$\tau=tr(\mathcal{A}_d)/tr(\mathcal{S}_d)$. The states are
conjectured to be bound entangled for $p_s<p\leq p_{1d}$.

\begin{figure}[ht]
\begin{center}
  \includegraphics [width=8 cm]{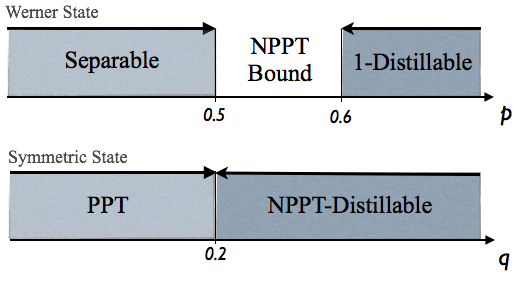}
\caption{{\footnotesize Entanglement properties of Werner,
$\rho_W$, and symmetric state, $\sigma(q)$, for the qutrit case
($d=3$). In the region between separability and 1-distillability,
$\rho _W$ is NPPT and conjectured bound. The point $q=0.2$
represents the extremal value for which states $\sigma(q)$ are
PPT, thus not distillable. For larger values of $q$ the states are
distillable (in particular, 1-distillable).}}\label{Wer&Activator}
\end{center}
\end{figure}

\textsc{Symmetric States.} Acting on an Hilbert space
$\mathcal{H}=\mathcal{H}_{A1} \otimes \mathcal{H}_{A2} \otimes
\mathcal{H}_{B1} \otimes \mathcal{H}_{B2}$, the symmetric states
under consideration commute with all unitaries of the form
$W=(U\otimes V)_A\otimes( U\otimes V^*)_B$ (where $V^*$ is the
complex conjugate of $V$). These states can be represented in a
compact form as~\cite{Symmetr}:
\begin{equation*}
\sigma= \sum_{i=1}^4 \lambda_i P_i/ tr[P_i]
\end{equation*}
where $P_1=\mathcal{A}_d^{(1)}\otimes \mathds{P}_d^{(2)}$,
$P_2=\mathcal{S}_d^{(1)}\otimes \mathds{P}_d^{(2)}$,
$P_3=\mathcal{A}_d^{(1)}\otimes \left(\mathds{1}-\mathds{P}_d
\right)^{(2)}$, $P_4=\mathcal{S}_d^{(1)}\otimes \left
(\mathds{1}-\mathds{P}_d \right)^{(2)}$. $\mathds{P}_d$  and
$\mathds{1}-\mathds{P}_d$ represent the projector onto the
maximally entangled state $|\psi_d ^+ \rangle =1/\sqrt{d}
\sum_{i=1}^{d}|ii\rangle $, and its orthogonal complement,
respectively. In Ref.~\cite{Activ} the authors identify a region
in the space of parameters $\lambda_i$ so that the state $\sigma$
(i) is bound entangled but (ii) gives a distillable state when
combined with a Werner state in the conjectured region of bound
entanglement. Among all the states with these properties, we focus
here on:
\begin{equation}\label{eq:symmetric}
\sigma (q)=q \frac{\mathcal{A}_d }{tr(\mathcal{A}_d)}\otimes \mathds{P}_d+(1-q) \frac{\mathcal{S}_d}{tr(\mathcal{S}_d)}\otimes  \frac{(\mathds{1}-\mathds{P}_d)}{tr(\mathds{1}-\mathds{P}_d)}
\end{equation}
where $q=1/(d+2)$. This state is a {\it universal activator}, in
the sense that it defines a distillable state when combined with
any entangled Werner state. It is also relevant for what follows
to study the distillability properties of
states~\eqref{eq:symmetric} for any value of $q$ and $d=3$. These
states are NPT and 1-distillable for $q>1/5$. The latter follows
from the fact that in this region, there exist local projections
on two-qubit subspaces mapping states~\eqref{eq:symmetric} onto an
entangled two-qubit state. The qubit subspaces are spanned by
$\ket{00},\ket{01}$ on Alice's side and $\ket{10},\ket{11}$ on
Bob's. Figure~\ref{Wer&Activator} summarizes the main entanglement
properties of these states.

\subsubsection{Protocol for Quantum Activation}

As already announced, any entangled Werner state, and in
particular any conjectured bound entangled Werner state, gives a
distillable state when combined with the universal activator
$\sigma(q)$ with $q=1/(d+2)$, simply denoted as $\sigma$. If
initially the two parties are sharing a Werner state $\rho$ acting
on $\mathcal{H}_0=\mathcal{H}_{A_0}\otimes \mathcal{H}_{B_0}$ and
a symmetric state $\sigma$ acting on
$\mathcal{H}_{1,2}=\mathcal{H}_{A_1} \otimes \mathcal{H}_{A_2}
\otimes \mathcal{H}_{B_1} \otimes \mathcal{H}_{B_2}$, each party
applies a projection onto a maximally entangled states on
$\mathcal{H}_{A_0}\otimes \mathcal{H}_{A_1}$ and
$\mathcal{H}_{B_0}\otimes \mathcal{H}_{B_1}$ respectively. The
resulting state is an isotropic state $\rho_{iso}$ acting on
$\mathcal{H}_{A_2} \otimes \mathcal{H}_{B_2}$. Recall that
isotropic state are $U\otimes U^*$ invariant and defined by the
convex combination of a maximally entangled state and white noise,
$\identity/d^2$. One can see that the resulting isotropic state
has an overlap with a maximally entangled state,
$\tr(\rho_{iso}\mathds{P}_d)$, larger than $1/d$ for any entangled
Werner state. As shown in~\cite{suffcond}, this condition is
sufficient for distillability.

\subsection{Classical Activation}
This section contains the main results of our work. Our goal is to
construct a classical cryptographic analog of the quantum
activation example discussed above.

We first associate probability distributions to all the previous
quantum states. In order to do so, we purify the initial bipartite
noisy quantum states $\rho_{AB}$ by including an environment, and
then map the tripartite quantum states $\ket{\psi_{ABE}}$ onto
probability distributions by performing some local measurements,
see~\eqref{pxyz}. The procedure to choose these measurements is
always the same: computational bases for the honest parties, and
general measurements for Eve. More precisely, denoting by $X$ and
$Y$ the result obtained by Alice and Bob, this
effectively projects Eve's system onto the pure state
$\ket{e_{XY}}=\langle XY\ket{\psi_{ABE}}$ with probability
$P(XY)=\bra{XY}\rho_{AB}\ket{XY}$. Given that, the measurement
that Eve applies is the one that minimizes her error probability
when distinguishing the states in the ensemble
$\{\ket{e_{XY}},P(XY)\}$. Note that this choice of measurement may
not necessarily be optimal from Eve's point of view in terms of
the secret correlations between Alice and Bob, but it seems a
natural choice. This procedure is applied to the two family of
states, namely Werner and symmetric. Because of the symmetries of
these states, the measurements minimizing Eve's error probability
can be analytically determined using the results
of Refs~\cite{Discrimin,SqMeas}.

In order to characterize the secrecy properties of the obtained
probability distributions, we compute the intrinsic information
when numerically possible and use AD protocols for distillability. We stress that the considered protocols distill a
secret key in the same region of parameters in which entanglement
distillation was possible for the initial quantum states. Finally,
we introduce a quantum-like activation protocol that maps the two
probability distributions into a new distribution in which Alice
and Bob each have a bit. We then prove that an AD protocol allows
distilling a secret key for some value of the parameters in which
the initial quantum states were non-distillable. However, we are
unable to close all the gap between entanglement and
1-distillability for the Werner state.

\subsubsection{Probability Distributions}
  \textsc{Werner states distribution.} We start by mapping the Werner states of
two qutrits onto a probability distribution $P_{XYZ}$ following
the recipe explained in the previous section. In this way, we get
a one-parameter family of probability distributions $P_{XYZ}$,
(see Table~\ref{TablePxyx} for details), which depends just on the
same parameter $p$ defining the initial Werner state
(\ref{eq:wernerP}). The resulting distributions are given in
Table~\ref{TablePxyx}. The indices for Eve's symbols specify her
guess on Alice's and Bob's symbols or, in other words, if Eve
outcome is $Z=z_{ij}$, the most probable outcomes for Alice and Bob
are $X=i$ and $Y=j$.

\begin{table}[htdp]
\begin{center}
\begin{equation*}
\setlength{\extrarowheight}{1.5pt}
\begin{array}{c||c|c|c|}

 & 0 & 1 & 2  \bigstrut[b] \\  \hline \hline
0 &  \lambda_1\quad (z_{00}) &
\frac{\lambda_1+\lambda_2}{2}\begin{cases}
\delta_Z & \textrm{$(z_{10})$}\\
1- \delta_Z & \textrm{$(z_{01})$}
\end{cases} & \frac{\lambda_1+\lambda_2}{2}\begin{cases}
\delta_Z & \textrm{$(z_{20})$}\\
1- \delta_Z  & \textrm{$(z_{02})$}
\end{cases}  \bigstrut \\ \hline

1 & \frac{\lambda_1+\lambda_2}{2}\begin{cases}
\delta_Z & \textrm{$(z_{01})$}\\
1- \delta_Z  & \textrm{$(z_{10})$}
\end{cases} &\lambda_1 \quad (z_{11}) &  \frac{\lambda_1+\lambda_2}{2}\begin{cases}
\delta_Z & \textrm{$(z_{21})$}\\
1- \delta_Z  & \textrm{$(z_{12})$}
\end{cases} \bigstrut \\ \hline

2 & \frac{\lambda_1+\lambda_2}{2}\begin{cases}
\delta_Z & \textrm{$(z_{02})$}\\
1- \delta_Z & \textrm{$(z_{20})$}
\end{cases}  &  \frac{\lambda_1+\lambda_2}{2}\begin{cases}
\delta_Z & \textrm{$(z_{12})$}\\
1- \delta_Z  & \textrm{$(z_{21})$}
\end{cases} &\lambda_1 \quad (z_{22}) \bigstrut \\ \hline
\end{array}
\end{equation*}
\caption{{\footnotesize Tripartite probability distributions derived from Werner
states~(\ref{eq:wernerP}). The parameters in the table are as
follows: $\lambda_1=(1-p)/6$, $\lambda_2=p/3$ and
$\delta_Z=(\sqrt{\lambda_1}-\sqrt{\lambda_2})^2/(2(\lambda_1+\lambda_2))$.
Rows (columns) represent Alice's (Bob's) symbols. Eve's symbols
are shown in parenthesis. For example, the cell $(X=0,Y=1)$ shows
that whenever Alice and Bob get (0,1) (which happen with
probability $(\lambda_1+\lambda_2)/2$), Eve correctly guesses the
symbol $z_{01}$ with probability $1-\delta_Z$, and makes an error
(symbol $z_{10}$) with probability $\delta_Z$. }} \label{TablePxyx}
\end{center}
\end{table}%

As done for entanglement, we now characterize these distributions
in terms of their secret correlations. Recall that for the
quantum case and qutrits, the state was entangled for $p>p_s=1/2$
and conjectured non-distillable for $p\leq p_{1d}=3/5$. As we show
next, the same values appear for the analogous classical
distributions. Concerning the point $p_s$,  we compute the
intrinsic information of the distributions in
Table~\ref{TablePxyx} by numerical optimization over all possible
channels by Eve. Of course, one can never exclude the existence of
local minima and, therefore, that the intrinsic information is
strictly smaller than what numerically obtained. One may wonder
why this computation is necessary. For instance, at the point
$p=p_s$ the quantum state is separable and, then, it is known that
there exists a measurement by Eve such that the intrinsic
information between Alice and Bob is zero for all measurements.
Note however that in terms of intrinsic information, the optimal
measurement by Eve is the one that prepares on Alice and Bob the
ensemble of product states compatible with the separable state
Alice and Bob share. This measurement is not necessarily the same
as the one minimizing Eve's error probability when Alice and Bob
measure in the computational bases. The same applies to the
entanglement region. While there are measurements such that Alice
and Bob share secret correlations no matter which measurement Eve
performs, these measurements are not on the computational bases.

\begin{figure}[htdp]
\begin{center}
  \includegraphics[width=15 cm]{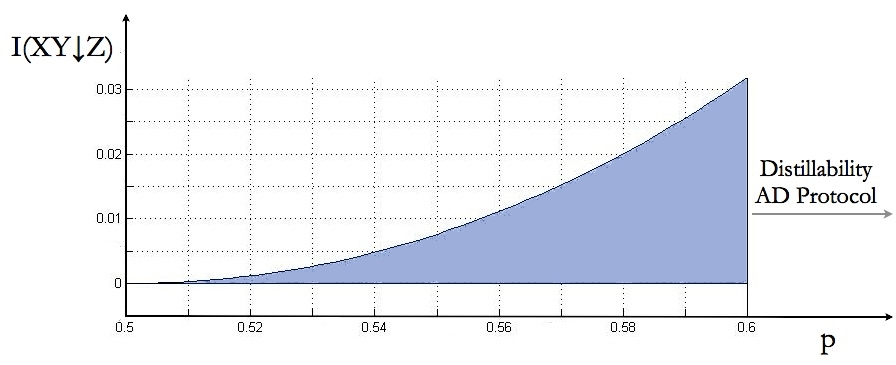}\\
  \caption{{\footnotesize Behaviour of the intrinsic information for the $P_{XYZ}$
  relative to the Werner state. Note that: i) $I(X;Y\downarrow Z)$
  is equal to $0$ at point $p=0.5$ which corresponds to the last
  point of separability for the Werner state; ii) $I(X;Y\downarrow Z)$
  is strictly positive at point $p=0.6$ which corresponds to
  the extreme value of $p$ for which it is 1-copy distillable.
  }}
  \label{Intr}
  \end{center}
\end{figure}

Using the numerical insight, we find a conjectured optimal channel
that reproduces the numerical results. The optimal channel gives
zero intrinsic information exactly at the point $p=p_s$. It
maps Eve's symbols $z_{ii}$ onto $z{ij}$ with $i\neq j$ with
equal probability ($i,j=0,1,2$). Its easy form leads to the
following analytical expression for $I(X;Y \downarrow Z)$:
\begin{eqnarray}
 \nonumber I(X;Y \downarrow Z) =
 -\log(1-x^2)-x\log\left(\frac{1+x}{1-x}\sqrt{\frac{\tau-2x}{\tau+2x}}\right)+\frac{\tau}{4}\log{(\tau^2-4x^2)}  +\left(1-\frac{\tau}{2}\right)\log(2-\tau)
\end{eqnarray}
where $\tau=1+p$, $x=\sqrt{2p(1-p)}$. Figure
\ref{Intr} shows the behavior  of this quantity in the region of
interest.

\vspace{1cm}

Moving to the distillability properties, we study AD protocols and
identify a value of $p$ for which positive secret-key rate can be
obtained by the two honest parties through these protocols. The
considered protocol is the quantum analogue of the quantum one and
uses a binaryzation. Alice and Bob first discard one (but the
same) of their symbols. Then, one of the parties, say Bob, applies
a local permutation to his symbols. For example, if they agreed on
discarding symbol 2, then Bob applies $0 \leftrightarrow 1$. Alice
and Bob now apply AD to the resulting two-bit distribution. This
distribution is shown in Table~\ref{TablePxyxbin}.

From the obtained table, it is possible to estimate the dependence
of Bob's and Eve's errors on the size of the blocks used for AD,
denoted by $N$. Recall that in the case of bits the protocols
works as follows: Alice generates a random bit $\zeta$ and chooses
$N$ symbols $\textbf a$ from her list of data. She then sends to
Bob the information about these symbols and the vector
$\textbf{\bar a}$ such that $a_i\oplus \bar a_i=\zeta,\forall i$.
Bob takes the symbols in his list corresponding to those chosen by
Alice, $\textbf b$, and accepts only when $\chi=b_i\oplus \bar
a_i,\forall i$. Bob's error probability $\beta_N$ is now easy to
compute. Denote by $\beta$ the error probability in the initial
two-bit probability distribution, $\beta=P(X\neq
Y)=2\lambda_1/(3\lambda_1+\lambda_2)$. Bob accepts a bit whenever
either all his $N$ symbols are identical to those of Alice, which
happens with probability $(1-\beta)^{N}$, or all his symbols are
different, whose probability is $\beta^{N}$. Thus, the probability
of accepting a wrong bit conditioned on acceptance is given by:
\begin{equation}\label{BobErr}
\beta_N=\frac{\beta^N}{\beta^N+(1-\beta)^N}\leqslant
\left(\frac{\beta}{1-\beta}\right)^N .
\end{equation}
The upper bound becomes tight in the limit $N\rightarrow\infty$.

\begin{table}[htdp]
\begin{center}
\begin{equation*}
\setlength{\extrarowheight}{1.5pt}
\begin{array}{c||c|c|}

 & 0 & 1   \bigstrut[b] \\  \hline \hline
0 & \frac{\lambda_1+\lambda_2}{2}\begin{cases}
\delta_Z & \textrm{$(z_{11})$}\\
1- \delta_Z & \textrm{$(z_{00})$}
\end{cases} & \lambda_1\quad (z_{01}) \bigstrut \\ \hline

1  &\lambda_1 \quad (z_{10}) &
\frac{\lambda_1+\lambda_2}{2}\begin{cases}
\delta_Z & \textrm{$(z_{00})$}\\
1- \delta_Z  & \textrm{$(z_{11})$}
\end{cases} \bigstrut \\ \hline

\end{array}
\end{equation*}
\caption{{\footnotesize Two-bit distribution resulting from
projecting the initial distribution of Table~\ref{TablePxyx} on
the space $X,Y=0,1$ and after Bob permutes his symbol. For the
sake of clarity, we apply a permutation also on the second index
of Eve's symbols, that is $z_{ij}\rightarrow z_{i1-j}$. All the
terms in the table should be normalized by a factor
$3\lambda_1+\lambda_2$.}} \label{TablePxyxbin}
\end{center}
\end{table}%

We now move to the estimation of Eve's error $\epsilon_N$. As her
information is probabilistic, there is always a non-zero
probability that she makes a mistake. For the estimation we
compute a lower bound on the error given by all the cases in which
the $N$ symbols observed by Eve do not provide her any information
about the value of the bit generated by Alice. In the computation,
it is simpler to use Eve's probabilities conditioned on the fact
that Alice and Bob have made no mistake after AD (which means that
no mistake has occurred for any of the $N$ symbols). Or in other
words, we only consider the terms in the diagonal of
Table~\ref{TablePxyxbin}. This does not make any difference for
what follows as in the limit $N\rightarrow\infty$ the probability
of Bob accepting a wrong symbol goes to zero. After Bob's
acceptance, Eve knows that the actual string $\textbf a$ used by
Alice is either equal to $\bar{\textbf{a}}$ (the one sent on the
public channel) when $\zeta=0$, or $\textbf{\bar a}'$ (the
permuted one, that is, $\bar a'_i=1-\bar a_i$) when $\zeta=1$.
Clearly, all the events in which the $N$ symbols observed by Eve,
$Z^{(i)}$, are such that $P(Z^{(1)}.. \ Z^{(N)}|\textbf
a=\bar{\textbf{a}})=$ $P(Z^{(1)}.. \ Z^{(N)}|\textbf
a=\textbf{\bar a}')$ do not give her any information about
$\zeta$. In these cases, Eve has to randomly guess Alice's symbol
and makes an error with probability $1/2$. Due to the symmetry in
the diagonal of Table~\ref{TablePxyxbin}, that is, $
P(Z=z_{00}|X=0)=P(Z=z_{11}=1|X=1)$ and
$P(Z=z_{11}|X=0)=P(Z=z_{00}|X=1)$, all the events where Eve has
exactly $N/2$ of her symbols equal to $z_{00}$ and $N/2$ equal to
$z_{11}$ satisfy the previous condition and, thus, contribute to
her error. Counting all the possible ways of distributing these
cases leads to the following lower bound on Eve's error
probability~\cite{QuantVsClass}:
\begin{equation}\label{EveErr}
\epsilon_N \geqslant \frac{1}{2}\left (
\begin{array}{c}
N \\
N/2
\end{array} \right )
  \delta_Z^{N/2}(1-\delta_Z)^{N/2}
\end{equation}
where $\delta_Z$ is the probability for Eve to guess wrongly
conditioned on those cases in which Alice and Bob's symbols
coincide (this value is made explicit in the caption of Figure
\ref{TablePxyx}). The asymptotic behavior of (\ref{EveErr}), after
applying the Stirling's approximation $(n!)^2\simeq(2n)!/2^{2n}$
and expanding the binomial coefficient can be expressed as:
\begin{equation}\label{EveErrS}
\epsilon_N \geqslant c (2\sqrt{\delta_Z (1-\delta_Z)})^N ,
\end{equation}
with $c$ being a positive constant.

By comparing Eqs. (\ref{BobErr}) and (\ref{EveErrS}) one concludes
that whenever
\begin{equation}\label{AD2}
\frac{\beta}{1-\beta}< 2\sqrt{\delta_Z (1-\delta_Z)}
\end{equation}
key distillation is possible. This follows from the fact that, if
this condition holds, Bob's error is exponentially smaller than
Eve's with $N$. This in turn implies that it is possible to choose
a value of $N$ such that Alice-Bob mutual information is larger than
Alice-Eve and one-way distillation techniques can distill a secret
key. From~\eqref{EveErrS} one gets that AD works whenever $p>3/5$,
as for 1-distillability in the quantum case. Before concluding
this part, we would like to mention that the same range of
parameters for distillation is obtained if one applies the
generalized AD protocol of Ref.~\cite{AGS}.\\
\smallskip

\textsc{Symmetric states distribution.} We apply the same machinery to the
symmetric states $\sigma(q)$. Again, the symmetries of the states
allow the explicit computation of the measurement by Eve
minimizing her error probability for any value of $q$. The
obtained distributions, denoted by $Q_{X1, Y1, X2, Y2, \tilde{Z}}$, is
significantly more complex and shown in Appendix A. It consists of
two trits for Alice, $(X_1,X_2)$ and two trits for Bob,
$(X_2,Y_2)$, while Eve's variable can take $63$ possible
values. It is now much harder to estimate the secrecy properties
of the distribution. For instance, we did not make any attempt to
compute the intrinsic information. However, we are able to show
that Alice and Bob can distill a secret key whenever $q>1/5$ as in
the quantum regime.

To simplify our task, we exploit again the concept of
\textit{binaryzation}. Inspired by the quantum projections used
for the distillation of $\sigma(q)$, Alice and Bob select two
outcomes on each side, namely $00,01$ for Alice and $10,11$ for
Bob.  The obtained two-bit distribution is shown in Table~\ref{TableQSmall}.

\begin{table}[htdp]
\begin{center}
\begin{equation*}
\setlength{\extrarowheight}{1.7pt}
\begin{array}{c||c|c|}
     & 0\: [10] & 1\: [11]  \bigstrut[b] \\  \hline \hline
0\: [00] & \frac{1+7q}{5+11q}
\begin{cases}
P_G & \textrm{$(\tilde{z}_{0100})$}\\
P_L & \textrm{$(\tilde{z}_{0111})$}\\
P_L & \textrm{$(\tilde{z}_{0122})$}\\
P_B & \textrm{$(\tilde{z}_{1000})$}\\
P_H & \textrm{$(\tilde{z}_{1011})$}\\
P_H & \textrm{$(\tilde{z}_{1022})$}\\
\end{cases} & \frac{3 (1-q)}{2 (5+11q)}\begin{cases}
1/2 & \textrm{$(\tilde{z}_{0101})$}\\
1/2  & \textrm{$(\tilde{z}_{1001})$}
\end{cases}  \bigstrut \\ \hline

1\: [01] &  \frac{3 (1-q)}{2 (5+11q)}\begin{cases}
1/2 & \textrm{$(\tilde{z}_{0110})$}\\
1/2  & \textrm{$(\tilde{z}_{1010})$}
\end{cases} &\frac{1+7q}{5+11q}
\begin{cases}
P_L & \textrm{$(\tilde{z}_{0100})$}\\
P_G & \textrm{$(\tilde{z}_{0111})$}\\
P_L & \textrm{$(\tilde{z}_{0122})$}\\
P_H & \textrm{$(\tilde{z}_{1000})$}\\
P_B & \textrm{$(\tilde{z}_{1011})$}\\
P_H & \textrm{$(\tilde{z}_{1022})$}\\
\end{cases}\bigstrut \\ \hline
\end{array}
\end{equation*}
\caption{{\footnotesize Two-bit distribution obtained as a result of the
binaryzation applied to $Q_{X1, Y1, X2, Y2, \tilde{Z}}$. Note that we
have relabeled the old symbols (shown in square brakets) by $0$
and $1$, in the following we use $\tilde{X}, \tilde{Y}$ to refer to them. The parameters in the table are as follows:
$\alpha=\sqrt{8q/(1+7q)}$ and $\gamma=\sqrt{(1-q)/(2 (1+7q))}$,
$P_G=(\alpha+2\gamma)^2/6$, $P_B=(-\alpha+2\gamma)^2/6$,
$P_L=(\alpha-\gamma)^2/6$, $P_H=(\alpha+\gamma)^2/6$. }}
\label{TableQSmall}
\end{center}
\end{table}%

They apply the standard bit AD protocol to this
distribution. As before, Bob's error can be easily computed,
getting the same as in Eq.~\eqref{BobErr}, but now with $\beta$ equal to $3 (1-q)/(5+11q)$. The estimation of Eve's error is much
more cumbersome. As above, the main idea is to derive a lower
bound on it based on those instances in which Eve's symbols do not
provide her any information about the symbol $\zeta$ Alice used
for AD. Again, one can restrict the analysis to the terms in the
diagonal of Table~\ref{TableQSmall}. The main difference in
comparison with the simple case discussed above is the larger
number of symbols for Eve. However, given the symmetry of the
distribution \ref{TableQSmall} it is enough to consider Eve's
symbols pair-wise:
\begin{gather*}
P(\tilde{Z}=\tilde{z}_{0100}|\tilde{X} \tilde{Y}=00)=P(\tilde{Z}=\tilde{z}_{0111}|\tilde{X} \tilde{Y}=11)=\bar\delta_1\\
P(\tilde{Z}=\tilde{z}_{0100}|\tilde{X} \tilde{Y}=11)=P(\tilde{Z}=\tilde{z}_{0111}|\tilde{X} \tilde{Y}=00)=\bar\eta_1\\
P(\tilde{Z}=\tilde{z}_{1000}|\tilde{X} \tilde{Y}=00)=P(\tilde{Z}=\tilde{z}_{1011}|\tilde{X} \tilde{Y}=11)=\bar\delta_2\\
P(\tilde{Z}=\tilde{z}_{1000}|\tilde{X} \tilde{Y}=11)=P(\tilde{Z}=\tilde{z}_{1011}|\tilde{X} \tilde{Y}=00)=\bar\eta_2\\
\end{gather*}
where we have used $\tilde{X}, \tilde{Y}$ to denote the
re-labeling of Alice and Bob's symbols. Note that the last two
subindexes of Eve's symbols are those that give her information
about Alice's (and Bob's) symbol. Symbols $\tilde{z}_{**22}$ give
her no information about Alice's symbols, so we sum them, their total probability being $\delta_3$. Given the public
string $\bar{\textbf{a}}_N$, one can see that all those cases for
which Eve has the same number $n_1$ of $\tilde{z}_{0100}$ and
$\tilde{z}_{0111}$ and the same number $n_2$ of $\tilde{z}_{1000}$
and $\tilde{z}_{1011}$, with $N=2n_1+2n_2+2n_3$ and where $2n_3$
is the total number of symbols $\tilde{z}_{**22}$, contribute to
her error. Thus, counting all these cases leads to the following
lower bound on Eve's error:

\begin{equation}\label{EveErrQsmall}
\epsilon_N \geqslant \frac{1}{2}\sum_{n_1,n_2,n_3}
\frac{N!}{(2n_1)!(2n_2)!(2n_3)!}
  \left( 2\sqrt{\delta_1 \eta_1} \right)^{2n_1}\left(2\sqrt{\delta_2 \eta_2} \right)^{2n_2} \left( \delta_3 \right )^{2n_3}
\end{equation}
where $\delta_i$ and $\eta_i$ are the probabilities shown above but normalized (since as already stated we are considering the asymptotic case).
After Stirling's approximation and summing eq. (\ref{EveErrQsmall}) the following compact form is obtained:
\begin{equation*}
\epsilon_N \geqslant c \left( 2\sqrt{\delta_1 \eta_1}+2\sqrt{\delta_2 \eta_2}+ \delta_3 \right)^N
\end{equation*}
with $c$ being a positive constant. Comparing the scaling of the
errors, one has that AD works whenever
\begin{equation}\label{AD3}
\frac{\beta}{1-\beta}< 2\sqrt{\delta_1 \eta_1}+2\sqrt{\delta_2 \eta_2}+ \delta_3
\end{equation}
where the right hand side is equal to $(\alpha+\gamma)^2/3$ (the
values of $\alpha$ and $\gamma$ are reported in the caption of
Table~\ref{TableQSmall}). Eq. (\ref{AD3}) is hence satisfied whenever
$q>\tilde{q}=0.2$, as announced.

\subsubsection{\normalsize{Protocol for Classsical Activation}}\label{QuanClassProt}
Inspired by the quantum activation example of Ref.~\cite{Activ},
we consider the following classical protocol. Alice and Bob have
access to the trits $X$ and $Y$, whose correlations are
described by $P_{XYZ}$, and the two trits $(X_1,X_2)$ and
$(Y_1,Y_2)$ correlated according to $Q_{X1, Y1, X2, Y2, \tilde{Z}}$.
Alice (Bob) keeps $X_2$ ($Y_2$), and only $X_2$ ($Y_2$), whenever
$X=X_1$ ($Y=Y_1$); otherwise they discard all the symbols.
This filtering projects the initial probability into a slightly
simpler two-trit distribution. The new probability distribution
$Q^*(X_2,Y_2, E)$ reads:
\begin{equation}
\label{eq:Q' } Q^*(X_2,Y_2, E) =\sum_{x,y=0}^2P(X=x,Y=y, Z) Q(X_1=x,Y_1=y,X_2,Y_2, \tilde{Z})
\end{equation}
where $E=[Z, \tilde{Z}]$ is the collection of Eve's ymbols. Finally Alice and Bob binaryze their
symbols by discarding one of the three values (the same for both),
say 2. The resulting distribution is shown in Table~\ref{Q*}.

As above, we use AD protocols to estimate the value of $p$ for
which Alice and Bob can extract a positive secret key rate if they
are sharing pairs of bits distributed according to Table~\ref{Q*}.
We are able to prove that whenever $p>p_c\simeq 0.513$ an AD
protocol allows distilling a secret key from the distribution in
Table~\ref{Q*} and, thus, a form of activation is possible.
Unfortunately, we are unable to reach the point $p=0.5$, as in the
quantum scenario. However, our analysis suggests that the secret
key rate is non-additive for some values of $p$. In the following
we summarize the key steps leading to this result.

\begin{figure}[htdp]
\begin{center}
  \includegraphics[width=11cm]{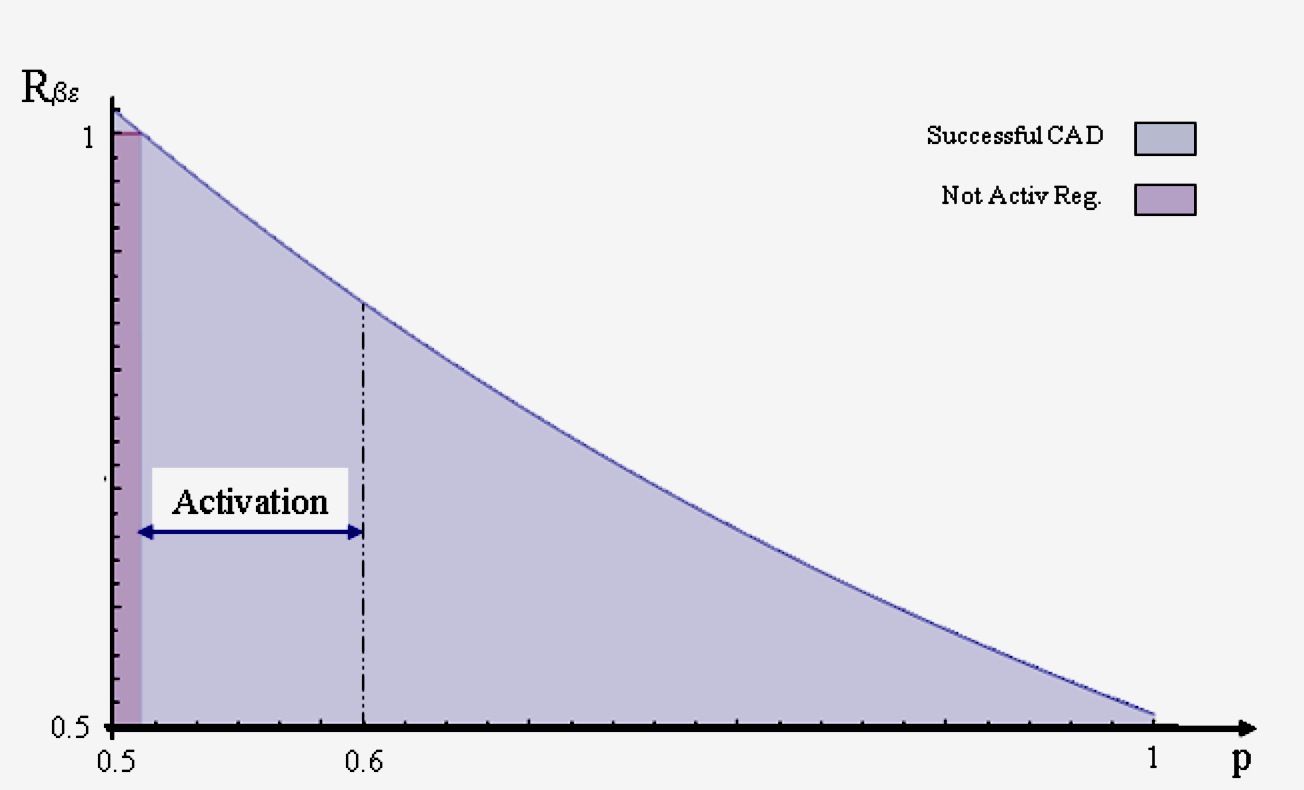}\\
  \caption{{\footnotesize The CAD protocol certifies that if the Werner state distribution (Table~\ref{TablePxyx}) is taken with $p>0.513$ positive secrecy can be extracted by the honest parties. Unfortunately, we cannot completely close the gap up to $p=0.5$. This would have shown a direct correspondence
  between the quantum and the classical scenario.}}\label{Activation}
\end{center}
\end{figure}

\begin{table}[htdp]
\setlength{\extrarowheight}{1.5pt}
\begin{equation*}
\begin{array}{c||c|c|}
& 0 & 1 \\
\hline \hline \multirow{11}{*}{0} & \hspace{0.35cm}
\frac{\lambda_1 (1-q)}{72 c_N}
\begin{footnotesize}
\begin{cases}
2/3\hspace{2.3cm}(z_{ii},\tilde{z}_{ii00})\\
1/6 \hspace{2.3cm}(z_{ii},\tilde{z}_{ii11}) \\
1/6\hspace{2.3cm}(z_{ii},\tilde{z}_{ii22})\\
\end{cases}\end{footnotesize}  &  \hspace{0.7cm}  \frac{\lambda_1 (1-q)}{48 c_N}
\begin{footnotesize}\begin{cases}
1 \hspace{1.3cm}(z_{00},\tilde{z}_{0001})\\
1 \hspace{1.3cm}(z_{11},\tilde{z}_{1101}) \\
1 \hspace{1.3cm}(z_{22},\tilde{z}_{2201})\\
\end{cases} \end{footnotesize}\\
 & \frac{(\lambda_1+\lambda_2)s_N}{2}
\begin{footnotesize}\begin{cases}
\delta_Z P_G+(1-\delta_Z )P_B  \hspace{0.1cm}(z_{ts},\tilde{z}_{st00})\\
\delta_Z P_L+(1-\delta_Z )P_H  \hspace{0.1cm}(z_{ts},\tilde{z}_{st11})\\
\delta_Z P_L+(1-\delta_Z )P_H  \hspace{0.1cm}(z_{ts},\tilde{z}_{st22})\\
\delta_Z P_B+(1-\delta_Z )P_G  \hspace{0.1cm}(z_{ts},\tilde{z}_{ts00})\\
\delta_Z P_H+(1-\delta_Z )P_L  \hspace{0.1cm}(z_{ts},\tilde{z}_{ts11})\\
\delta_Z P_H+(1-\delta_Z )P_L  \hspace{0.1cm}(z_{ts},\tilde{z}_{ts22})\\
\delta_Z P_B+(1-\delta_Z)P_G  \hspace{0.1cm}(z_{st},\tilde{z}_{st00})\\
\delta_Z P_H+(1-\delta_Z)P_L  \hspace{0.1cm}(z_{st},\tilde{z}_{st11}) \\
\delta_Z P_H +(1-\delta_Z)P_L \hspace{0.1cm}(z_{st},\tilde{z}_{st22})\\
\delta_Z P_G+(1-\delta_Z)P_B   \hspace{0.1cm}(z_{st},\tilde{z}_{ts00}) \\
\delta_Z P_L+(1-\delta_Z )P_H  \hspace{0.1cm}(z_{st},\tilde{z}_{ts11})\\
\delta_Z P_L+(1-\delta_Z) P_H  \hspace{0.1cm}(z_{st},\tilde{z}_{ts22})\\
\end{cases}\end{footnotesize}
  & \frac{( \lambda_1+\lambda_2)(1-q)}{192 c_N}
\begin{footnotesize}\begin{cases}
1/2  \hspace{1cm}(z_{01},\tilde{z}_{0101})\\
1/2  \hspace{1cm}(z_{01},\tilde{z}_{1001})\\
1/2  \hspace{1cm}(z_{10},\tilde{z}_{0101})\\
1/2  \hspace{1cm}(z_{10},\tilde{z}_{1001})\\
1/2  \hspace{1cm}(z_{02},\tilde{z}_{0201})\\
1/2  \hspace{1cm}(z_{02},\tilde{z}_{2001})\\
1/2  \hspace{1cm}(z_{20},\tilde{z}_{0201}) \\
1/2  \hspace{1cm}(z_{20},\tilde{z}_{2001})\\
1/2  \hspace{1cm}(z_{12},\tilde{z}_{1201}) \\
1/2  \hspace{1cm}(z_{12},\tilde{z}_{2101})\\
1/2  \hspace{1cm}(z_{21},\tilde{z}_{1201})\\
1/2  \hspace{1cm}(z_{21},\tilde{z}_{2101})\\
\end{cases}\end{footnotesize}\\
\hline \multirow{11}{*}{1} & \hspace{0.7cm}\frac{\lambda_1
(1-q)}{48 c_N}
\begin{footnotesize}
\begin{cases}
1 \hspace{1.3cm}(z_{00},\tilde{z}_{0010})\\
1 \hspace{1.3cm}(z_{11},\tilde{z}_{1110}) \\
1 \hspace{1.3cm}(z_{22},\tilde{z}_{2210})\\
\end{cases}\end{footnotesize} &
\hspace{0.3cm} \frac{\lambda_1 (1-q)}{72 c_N}
\begin{footnotesize}
\begin{cases}
1/6 \hspace{2.3cm}(z_{ii},\tilde{z}_{ii00})\\
2/3 \hspace{2.3cm}(z_{ii},\tilde{z}_{ii11}) \\
1/6 \hspace{2.3cm}(z_{ii},\tilde{z}_{ii22})\\
\end{cases}\end{footnotesize} \\
&
 \frac{( \lambda_1+\lambda_2)(1-q)}{192 c_N}
\begin{footnotesize}\begin{cases}
1/2  \hspace{1cm}(z_{01},\tilde{z}_{0110})\\
1/2  \hspace{1cm}(z_{01},\tilde{z}_{1010})\\
1/2  \hspace{1cm}(z_{10},\tilde{z}_{0110})\\
1/2  \hspace{1cm}(z_{10},\tilde{z}_{1010})\\
1/2  \hspace{1cm}(z_{02},\tilde{z}_{0210})\\
1/2  \hspace{1cm}(z_{02},\tilde{z}_{2010})\\
1/2  \hspace{1cm}(z_{20},\tilde{z}_{0210}) \\
1/2  \hspace{1cm}(z_{20},\tilde{z}_{2010})\\
1/2  \hspace{1cm}(z_{12},\tilde{z}_{1210}) \\
1/2  \hspace{1cm}(z_{12},\tilde{z}_{2110})\\
1/2  \hspace{1cm}(z_{21},\tilde{z}_{1210})\\
1/2  \hspace{1cm}(z_{21},\tilde{z}_{2110})\\
\end{cases}\end{footnotesize} &
\frac{(\lambda_1+\lambda_2)s_N}{2}
\begin{footnotesize}\begin{cases}
\delta_Z P_L+(1-\delta_Z )P_H  \hspace{0.1cm}(z_{ts},\tilde{z}_{st00})\\
\delta_Z P_G+(1-\delta_Z )P_B  \hspace{0.1cm}(z_{ts},\tilde{z}_{st11})\\
\delta_Z P_L+(1-\delta_Z )P_H  \hspace{0.1cm}(z_{ts},\tilde{z}_{st22})\\
\delta_Z P_H+(1-\delta_Z )P_L  \hspace{0.1cm}(z_{ts},\tilde{z}_{ts00})\\
\delta_Z P_B+(1-\delta_Z )P_G  \hspace{0.1cm}(z_{ts},\tilde{z}_{ts11})\\
\delta_Z P_H+(1-\delta_Z )P_L  \hspace{0.1cm}(z_{ts},\tilde{z}_{ts22})\\
\delta_Z P_H+(1-\delta_Z) P_L  \hspace{0.1cm}(z_{st},\tilde{z}_{st00})\\
\delta_Z P_B+(1-\delta_Z) P_G  \hspace{0.1cm}(z_{st},\tilde{z}_{st11}) \\
\delta_Z P_H +(1-\delta_Z)P_L \hspace{0.1cm}(z_{st},\tilde{z}_{st22})\\
\delta_Z P_L+(1-\delta_Z) P_H  \hspace{0.1cm}(z_{st},\tilde{z}_{ts00}) \\
\delta_Z P_G+(1-\delta_Z )P_B  \hspace{0.1cm}(z_{st},\tilde{z}_{ts11})\\
\delta_Z P_L+(1-\delta_Z) P_H  \hspace{0.1cm}(z_{st},\tilde{z}_{ts22})\\
\end{cases}\end{footnotesize}\\ \hline
\end{array}
\end{equation*}
\caption{{\footnotesize Resulting tripartite distribution after
the application of the classical protocol by the two honest
parties. The initial probability distributions $P_{XYZ}$ and
$Q_{X1, Y1, X2, Y2, \tilde{Z}}$ are mapped to the new
probability distribution $Q^*(X_2, Y_2, E)$ shown above. From this
classical object we can derive the minimum value of $p$ for which
positive secret key can be extracted by A and B. The parameters
that appear above are expressed as a function of p and q, the two
key parameters in the initial probability distributions.
$c_N=(\lambda_1+\lambda_2)(5+11q)/48+5 \lambda_1(1-q)/24$,
$s_N=(1+7q)/(144 c_N)$, $i,s,t=0,1,2$ with $s\neq t$ and $s<t$. In
our procedure the optimal $q$ for the symmetric state distribution
is taken equal to $1/5$.}} \label{Q*}
\end{table}
As mentioned, the values of interest for $P_{XYZ}$ and $Q_{X1,
Y1, X2, Y2, \tilde{Z}}$ are, $0.5 < p \leqslant 0.6 $ and
$q=0.2$, respectively. The distribution $Q^*(X_2, Y_2, E)$
resulting from the local filtering by the honest parties depends
on the parameter $p$. In order to estimate Eve's error we follow a
similar argument as for $Q_{X1, Y1, X2, Y2, \tilde{Z}}$, now
adapted to this slightly more complex case. Despite the big amount
of symbols on Eve's side (see Table~\ref{Q*}), the symmetry in the
distribution leads to six main classes that are relevant for the
AD analysis (appendix B further clarifies this point). These
arguments lead to the following bound on Eve's error:

\begin{equation}\label{EveErrFin}
\epsilon_N \geqslant \frac{1}{2}\sum_{n_1,n_2\ldots n_6}
\frac{N!}{(2n_1)! \ldots (2n_6)!}
  \left(6\sqrt{\delta_{1}\ \eta_{1}}\ \right)^{2n_1}\dots \left(6\sqrt{\delta_{5}\ \eta_{5}}\right)^{2n_5} \delta_6^{2n_6}
\end{equation}
where $\sum^{6}_{i=1} 2n_i=N$. Note that as before the terms
$\delta_i \eta_i$ with $i=1\ldots 5$ take into account those cases
in which Eve has $n_i$ symbols that coincide with the public
string sent by Alice and $n_i$ symbols that are opposite to those
appearing in the public string. The last term, $\delta_6$, as
before, refers to the sum of probabilities for which Eve has no
information at all (see details in appendix B). In the asymptotic
case we are treating here, Eq. (\ref{EveErrFin}) converges to a
multinomial distribution, namely:
\begin{equation}\label{EEFA}
\epsilon_N \geqslant c
 \left( 6\left(\sqrt{\delta_{1}\ \eta_{1}}\ + \dots \sqrt{\delta_{5}\
 \eta_{5}}\right)+\delta_6\right)^N
\end{equation}
with $c$ being a positive constant. Bob's error is much easier to
compute, getting $\beta=(3\lambda_1+\lambda_2)(1-q)/(16c_N)$.
Putting these two terms together, we have that the AD protocols
works whenever:
\begin{equation}
\frac{\beta}{1-\beta}< 6\left(\sqrt{\delta_{1}\ \eta_{1}}\ + \dots \sqrt{\delta_{5}\
 \eta_{5}}\right)+\delta_6
\end{equation}
Figure \ref{Activation} shows the ratio between the left hand side
and the right hand side, $R_{\beta \epsilon}$, as a function of
the parameter $p$. As above, whenever $R_{\beta\epsilon}<1$, the
AD protocol succeeds. The point at which $R_{\beta\epsilon}=1$
corresponds to $p=0.513$, as already announced.

\section{Conclusions}
\label{Conclusions}

Non-additivity is an ubiquitous phenomenon in Quantum Information
Theory due to the presence of entanglement. In this work, we
provide some evidence for the existence of similar effects for
secret classical correlations. Exploiting the analogies between
the entanglement and secret-key agreement scenario, we have shown
that two classical distributions from which no secrecy can be
extracted by AD protocols can lead to a positive secret key rate
when combined.

The evidence we provide is somehow similar to the conjectured
example of activation for bipartite entangled states. Note however
that, in the quantum case, one of the two states is provably
bound. As mentioned several times, it could well happen that one,
or even the two probability distributions considered here are
key-distillable. Indeed, there exist examples of bound entangled
states from which one can obtain probability distributions with
positive secret-key rate~\cite{pbits}. Note however that all the
known examples of bound entangled states with non-zero privacy are
based on the existence of ancillary systems on the honest parties,
known as shields, that prevent Eve from having the purification of
the systems Alice and Bob measure to construct the key. If any of
the probability distributions constructed here were key
distillable, they would constitute a novel example of secret
correlations from a bound entangled state that does not fit in the
construction of~\cite{pbits}.

\subsection*{Acknowledgement}
We thank Lluis Masanes for contributions at early stages of this project. This work was supported by the ERC starting grant PERCENT, the European EU FP7 Q-Essence and QCS projects, the Spanish FIS2010-14830, Consolider-Ingenio QOIT and Chist-Era DIQIP projects.

\newpage
\appendix
\section*{\normalsize{Appendix A}}
This appendix shows the probability distribution obtained by Alice, Bob and Eve after measuring the symmetric state (\ref{eq:symmetric}). Being the table very big we try to give here a schematic representation of it which can be equivalently useful to the reader to follow our arguments. It reads:

\begin{table}[htdp]
\begin{center}
\begin{equation*}
\setlength{\extrarowheight}{1.2 pt}
\begin{array}{c||c|c|c||c|c|c||c|c|c|}

      &    00    &      01 & 02     & 10 & 11 & 12 & 20 & 21 & 22  \bigstrut[v] \\  \hline \hline
 00 & (1_u) &  + & + & (2_u) & * & * & (2_w) & * & * \\ \hline
 01 &  +  & (1_u) & + & * & (2_u) & * & * & (2_w)& * \\ \hline
 02 & + & + & (1_u) & * & * & (2_u) & * & * & (2_w) \\ \hline \hline
 10 & (2_u) & * & * & (1_v) & + & + & (2_v) & * & * \\ \hline
 11 & * & (2_u) & * & + & (1_v) & + & * & (2_v) & *\\ \hline
 12 & * & * & (2_u) & + & +  & (1_v) & * & * & (2_v)  \\ \hline \hline
 20 & (2_w) & * & * & (2_v) & * & * & (1_w) & + & + \\ \hline
 21 & * & (2_w) & * & * & (2_v) & * & + &  (1_w) & + \\ \hline
 22 & * & * & (2_w) & * & * & (2_v) & + & + & (1_w) \bigstrut \\ \hline
\end{array}
\end{equation*}
\end{center}
\caption{{\footnotesize Schematic view of the distribution $Q_{X1, Y1, X2, Y2,
\tilde{Z}}$. Due to the lack of space, cells have been grouped in terms of probability distributions and number of elements (symbols) as explained below.}} \label{TableQxyx}
\end{table}
\noindent The joint probabilities $P(X_1=i, Y_1=k,  X_2=j, Y_2=l)$ between the
honest parties are distributed as follows:
\begin{itemize}
\item[-] cells of type $(1_i)$, with $i=u,v,w$ are equal to $\frac{1-q}{72}$
\item[-] cells of type $(2_i)$, with $i=u,v,w$,  are equal to $\frac{1+ 7q}{144}$;
\item[-] cells of type $*$ ,  are equal to $\frac{1-q}{48}$;
\item[-] cells of type $+$ ,  are equal to $\frac{1-q}{96}$;
\end{itemize}
Concerning Eve's side (see caption of Table~\ref{TablePxyx} for
more details about how to read the tables):
\begin{itemize}
\item[-] cells of type $(1_i)$, with $i=u,v,w$ contain three elements. The terms that play a role in her discrimination are indicated by the same number and subindex letter. For example, consider the cell $X_1=0, Y_1=0,  X_2=0, Y_2=0$. The label  $1_u$ is used for this cell (the same one indicates $X_1=0, Y_1=0,X_2=1, Y_2=1$ and $X_1=0, Y_1=0,X_2=2, Y_2=2$). The three elements here are the three probability distributions:
\begin{center}
$P(0,0,0,0,\bar{z}_{00,00}), \quad P(0,0,0,0,\bar{z}_{00,11}), \quad
P(0,0,0,0,\bar{z}_{00,22}).$
\end{center}
$P(0,0,0,0,\bar{z}_{00,00})$ refers to the probability that Eve guesses
correctly, the remaining two $P(0,0,0,0,\bar{z}_{00,11})$,
$P(0,0,0,0,\bar{z}_{00,22})$ refers to the probability she guesses wrongly.
\item[-] cells of type $(2_i)$, with $i=u,v,w$,  contain six elements;
\item[-] cells of type $*$ , contain two elements distributed with probability one half (in this cases, she knows nothing about A and B symbols) ;
\item[-] cells of type $+$ , contains only one term since in this case Eve's symbol is perfectly correlated with those of A and B;
\end{itemize}

\section*{\normalsize{Appendix B}}
In this second appendix, we clarify why it is enough to consider six classes of distributions in the AD analysis of section~\ref{QuanClassProt}.
From Table~\ref{Q*} the following relations hold:
\begin{gather*}
P(E=[z_{ii},\tilde{z}_{ii00}]|X_2 Y_2=00)=P(E=[z_{ii},\tilde{z}_{ii11}]|X_2 Y_2=11)=\bar\delta_1\\
P(E=[z_{ii},\tilde{z}_{ii11}]|X_2 Y_2=00)=P(E=[z_{ii},\tilde{z}_{ii00}]|X_2 Y_2=11)=\bar\eta_1\\
P(E=[z_{ts},\tilde{z}_{st00}]|X_2 Y_2=00)=P(E=[z_{ts},\tilde{z}_{st11}]|X_2 Y_2=11)=\bar\delta_2\\
P(E=[z_{ts},\tilde{z}_{st11}]|X_2 Y_2=00)=P(E=[z_{ts},\tilde{z}_{st00}]|X_2 Y_2=11)=\bar\eta_2\\
P(E=[z_{ts},\tilde{z}_{ts00}]|X_2 Y_2=00)=P(E=[z_{ts},\tilde{z}_{ts11}]|X_2 Y_2=11)=\bar\delta_3\\
P(E=[z_{ts},\tilde{z}_{ts11}]|X_2 Y_2=00)=P(E=[z_{ts},\tilde{z}_{ts00}]|X_2 Y_2=11)=\bar\eta_3\\
P(E=[z_{st},\tilde{z}_{st00}]|X_2 Y_2=00)=P(E=[z_{st},\tilde{z}_{st11}]|X_2 Y_2=11)=\bar\delta_4\\
P(E=[z_{st},\tilde{z}_{st11}]|X_2 Y_2=00)=P(E=[z_{st},\tilde{z}_{st00}]|X_2 Y_2=11)=\bar\eta_4\\
P(E=[z_{st},\tilde{z}_{ts00}]|X_2 Y_2=00)=P(E=[z_{st},\tilde{z}_{ts11}]|X_2 Y_2=11)=\bar\delta_5\\
P(E=[z_{st},\tilde{z}_{ts11}]|X_2 Y_2=00)=P(E=[z_{st},\tilde{z}_{ts00}]|X_2 Y_2=11)=\bar\eta_5\\
\end{gather*}
and $\bar\delta_6$ is the sum of all the $P(E=[z_{**},\tilde{z}_{**22}]|X_2=Y_2)$. As already stated in the caption of Table~\ref{Q*}, $i,s,t=0,1,2$ with
$s\neq t$ and $s<t$.  In the computation, it is simpler
to use Eve's probabilities conditioned on the fact that Alice and
Bob have made no mistake after AD, so this means that we only need to
consider the terms in the diagonal of Table~\ref{Q*}. For this reason the $\delta_i, \eta_i$ appearing in eq. (\ref{EveErrFin}) are the previous ones but normalized. The complete expression is then derived according to the argument already presented at page \pageref{EveErrQsmall}.

\end{document}